# A Grid Information Infrastructure for Medical Image Analysis

Dmitry Rogulin[1,2], Florida Estrella[1], Tamas Hauer[1,2], Richard McClatchey[1]
& Tony Solomonides[1]

[1] CCCS Research Centre, University of the West of England, Frenchay, Bristol BS16 1QY
[2] ETT Division, CERN, 1211 Geneva 23, Switzerland

**Abstract**

The storage and manipulation of digital images and the analysis of the information held in those images are essential requirements for next-generation medical information systems. The medical community has been exploring collaborative approaches for managing image data and exchanging knowledge and Grid technology [1] is a promising approach to enabling distributed analysis across medical institutions and for developing new collaborative and cooperative approaches for image analysis without the necessity for clinicians to co-locate. The EU-funded MammoGrid project [2] is one example of this and it aims to develop a Europe-wide database of mammograms to support effective co-working between healthcare professionals across the EU. The MammoGrid prototype comprises a high-quality clinician visualization workstation (for data acquisition and inspection), a DICOM-compliant interface to a set of medical services (annotation, security, image analysis, data storage and querying services) residing on a so-called 'Grid-box' and secure access to a network of other Grid-boxes connected through Grid middleware. One of the main deliverables of the project is a Grid-enabled infrastructure that manages federated mammogram databases across Europe. This paper outlines the MammoGrid Information Infrastructure (MII) for meta-data analysis and knowledge discovery in the medical imaging domain.

## 1. Medical Imaging Requirements

Medical diagnosis and intervention increasingly relies upon images, of which there is a growing range available to the clinician: X-ray (increasingly digital, though still overwhelmingly film-based), ultrasound, MRI, CT, PET scans etc. This reliance will increase as high bandwidth systems for picture archiving and communications are installed in large numbers of hospitals (currently, primarily in large teaching hospitals). Digital medical images represent significant amounts of information collected by healthcare institutions about patients and their medical conditions. Clinicians need to have (strictly regulated) access to medical image data in order to conduct the analyses that can lead to detection and prevention of diseases such as breast cancer. This implies creating diagnostic tools which allow efficient management of data, enhanced analysis, quality control [3] and acquisition procedures, epidemiological studies, education and knowledge exchange in the medical imaging domain.

However, there are a number of factors that make patient management based on medical images particularly difficult. First, there is a growing range of images available to clinicians, captured by different, potentially highly variable, medical devices and where the same image may look different depending on machine and parameters. Different technical and technological solutions may result in the same medical data being stored at different locations in different formats. This leads to significant heterogeneity in the nature and content of those images. Additional heterogeneity may be due to difference in data model, data semantics and schemata, etc.

Secondly, medical images that have been captured in different geographical locations may need to be distributed across many hospitals for analysis over the lifetime of a patient (i.e. patient data are not necessarily held at a single site or accessible through a common interface). This implies a natural distribution of medical data which is particularly acute when studies are required to be conducted across sets of images resident in different computer systems.

Thirdly, very large quantities of data (from many megabytes to several gigabytes) with very complex structures are involved in medical imaging processing and exchange. In addition large numbers of exemplars are required to include sufficient numbers of abnormal cases to make significant statistical inferences in image analysis. These cases may be obtained from many geographically remote locations often crossing national boundaries. These factors lead to a need for the sharing of computation and storage resources across hospitals so that medical image data can be studied not only for individuals but also for populations of patients and at a high degree of accuracy and confidentiality as well as ease and transparency of access to the medical data. This implies providing environments where people are connected through common technologies, where new partners can join or leave without needing a major change in the system, where data can be shared inside virtual communities or organisations that respect agreed access regulations and where different fields can start working together by making resources (for example databases) accessible to each other.

Finally, the highly confidential nature of medical data requires that patient privacy and data security are key issues in collaborative medical image analysis. Data protection, ownership, data security and medical anonymity are issues that have to be addressed by any deployable distributed image analysis solution. Security in grid infrastructures may be sufficient for research, but it must be improved in the future to ensure privacy of data. Encrypted transmission and storage is not sufficient, integrity of data and automatic pseudonymization or anonymization must be enforced to guarantee that data is complete and reliable and that privacy leakage cannot occur due to unauthorized use of the resources (see [4]).

In an effort to address some of these key requirements for medical imaging, this paper outlines the provision of a so-called 'Information Infrastructure' or enabling Grid-based image analysis platform for the MammoGrid [2] project and describes the architecture chosen in that project. In the next section the appropriateness of Grids is considered for image analysis before the MammoGrid project is described in more detail in the following section. The MammoGrid Information Infrastructure is outlined in section 4 and an example is given for its use in query handling – a crucial element in medical image analysis. Conclusions and a discussion of future work complete the content of this paper.

## 2. The Grid for Medical Image Analysis

In recent years the Grid has emerged as one flavour of large-scale distributed computing that can address the specific requirements emerging from High Energy Physics (HEP) for the exploitation of distributed computing power and the automatic access to distributed data storage. The Grid is defined as "flexible, secure, coordinated resource sharing among dynamic collections of individuals, institutions and resources"[5]. Geographically separated but working collectively together to solve a problem groups of people can be organized in collaborations (referred to as 'virtual organisations', VOs) and use the shared resources of the Grid.

The Grid can provide a virtual platform for large-scale, resource-intensive, and distributed applications. It offers a connectivity environment allowing management and coordination of diverse and dispersed resources. The Grid enables access to increased storage and computing capacity providing mechanisms for sharing and transferring large amounts of data as well as aggregating distributed resources for running computationally expensive procedures. Another important point is that the Grid utilizes a common infrastructure based on open standards thus providing a platform for interoperability and interfacing between different Grid-based applications from the particular domain.

Grid technology can potentially enable medical applications an architecture for easy and transparent access to distributed heterogeneous resources (like data storage, networks, computational resources) across different organizations and administrative domains. The Grid offers a configurable environment allowing grid structures to be reorganized dynamically without disturbing any overall active Grid processing. In particular the Grid can address some of the following issues relevant to medical domains:

- *Data distribution:* The Grid provides a connectivity environment for medical data distributed over different sites. It solves the transparency issue by providing mechanisms which permit seamless access to and the management of distributed data. These mechanisms include services which deal with virtualization of distributed data regardless of their location.

- *Heterogeneity:* The Grid addresses the issue of heterogeneity by developing common interfaces for access and integration of diverse data sources. Such generic interfaces for consistent access to existing, autonomously managed databases that are independent of underlying data models are defined by the Global Grid Forum Database Access and Integration Services (GGF-DAIS) [6] working group. These interfaces can be used to represent an abstract view of data sources which can permit homogeneous access to heterogeneous medical data sets.

- *Data processing and analysis:* The Grid offers a platform for transparent resource management in medical analysis. This allows the virtualization and sharing of all resources (e.g. computing resources, data storage, etc.) connected to the grid. For handling computationally intensive procedures (e.g. CADe [7]), the platform provides automatic resource allocation and scheduling and algorithm execution, depending on the availability, capacity and location of resources.

- *Security and confidentiality*: Enabling secure data exchange between hospitals distributed across networks is one of major concerns of medical applications. Grid addresses security issues by providing a common infrastructure (Grid Security Infrastructure) for secure access and communication between grid-connected sites. This infrastructure includes authentication and authorization mechanisms, among other things, supporting security across organizational boundaries.

- *Standardization and compliance:* Grid technologies are increasingly being based on a common set of open standards (such as XML, SOAP, WSDL, HTTP etc.) and this is promising for future medical image analysis standards.

Effective medical imaging tools should provide solutions that allow ease and transparent access to image data regardless of physical location and storage format. There are a number of projects investigating the use of the Grid in medical informatics. As one example, the next section discusses the MammoGrid approach for managing mammography data on the Grid.

## 3. The MammoGrid Project

The set of medical imaging requirements listed in the previous section has been identified across the 'Grids in Healthcare' community [4] and especially in the current EU Framework V MammoGrid project [2]. In MammoGrid there is a clear requirement for an infrastructure that enables the very large volume of data associated with regional breast cancer screening programmes to be available across multiple medical centres in an acceptable time. These images must be processed through standardization software to eliminate any biases introduced during image acquisition so that a comparison of images from different patients and centres can be undertaken. In addition the standardised images must be made available to a data-mining engine that enables queries based on patient details and text annotations to be resolved. Furthermore the images must also be accessible to analysis algorithms that provide quantitative information, which is otherwise unavailable from visual inspection alone, and to detection systems that help in visual diagnosis.

The MammoGrid project aims to deliver a Grid-enabled infrastructure which federates multiple mammogram databases. This permits clinicians to develop new common, collaborative and co-operative approaches to the analysis of mammography data. Using the MammoGrid they will be able to quickly harness the use of massive amounts of medical image data to perform epidemiological studies, advanced image processing, radiographic education and ultimately, telediagnosis over communities of medical 'virtual organisations' i.e. groups that are geographically separated but working together using the shared resources of the Grid. This is achieved through the use of Grid-compliant services for managing (versions of) massively distributed files of mammograms, for handling the distributed execution of mammograms analysis software, for the development of Grid-aware algorithms and for the sharing of resources between multiple collaborating medical centres. By moving towards a service-oriented architecture (SOA) and supplying services interfaced to a Grid hosting environment through an Open Grid Services Architecture (OSGA) interface, the project aims to provide a path towards future Grid compliance.

The MammoGrid project decided to adopt a lightweight Grid middleware solution (called AliEn (Alice Environment) [8]). AliEn is a Grid framework developed to satisfy the needs of the ALICE high energy physics experiment at CERN for large scale distributed computing. It is built on top of the latest Internet standards for information exchange and authentication (SOAP, SASL, PKI) and common Open Source components (such as Globus/GSI, OpenSSL, OpenLDAP, SOAPLite, MySQL). AliEn provides a virtual file catalogue that allows transparent access to distributed datasets and provides top to bottom implementation of a lightweight Grid applicable to cases where handling of a very large number of files is needed. MammoGrid aims to deliver a set of medical services through a series of prototypes following an SOA philosophy and with an OGSA interface so that when suitable OGSA-based Grid solutions become available the MammoGrid services would be able to interface with those hosting environments.

In the deployment phase, the CERN AliEn middleware has been installed and configured on a set of novel 'Gridboxes', secure hardware units which are meant to act as each hospital's single point of entry onto the MammoGrid. These units are configured and tested at CERN and Oxford, for later testing and integration with other Gridboxes at hospitals in Udine (Italy) and Cambridge (UK). As the MammoGrid project has developed and new layers of Grid functionality have became available, AliEn has facilitated the incorporation of new stable versions of Grid software in a manner that has catered for controlled system evolution and has provided a rapidly available lightweight but highly functional Grid architecture for MammoGrid. It is envisaged that the AliEn code will further evolve as part of the emerging new Grid middleware of the EGEE project [9], and that the MammoGrid application will thus experience a seamless transition between existing and future middleware platforms.

## 4. The MammoGrid Information Infrastructure architecture

### 4.1. The High-level MammoGrid architecture

The MammoGrid architecture, schematically shown on figure 1, is based on SOA with standardized interfaces to both client front-end (e.g. MammoGrid Workstation, acquisition systems, medical devices) and to Grid middleware. A set of MammoGrid services facilitate the exchange of medical data between a MammoGrid Workstation and the Grid middleware. It is composed of two service layers: medical imaging (MI) services and grid-aware (GA) services.

MI-services provide a generic framework for handling image-related data. These services are directly invoked by clients through MammoGrid API (Application Programming Interface) functions. These include DICOM services for parsing (e.g. DICOM-to-XML), updating (e.g. XML-to-DICOM), uploading/downloading DICOM files, Data Transformation services for translating medical data between different formats and representations, Query Translator service etc. GA-services serve as mediating components between MI-services and the underlying Grid middleware. These services control and coordinate the hospital interactions with the encoded information of the medical imaging domain and the Grid. They interact with Grid API services and promote loose coupling between the MI-domain and Grid middleware.

It is necessary to mention that the Grid middleware is treated as a 'black-box', i.e. all interactions between the MammoGrid services and the Grid are assured through the Grid API services provided by the underlying Grid middleware. Such Grid API services can expose functions for authentication, file catalogue and database management, job submission and execution etc. The MammoGrid API offers an extensible set of functions used by clients as a means for providing access to MammoGrid services. These functions represent workflows of typical interactions between the MammoGrid Workstation and the Grid (such as adding DICOM files, querying associated data, submitting and executing jobs).

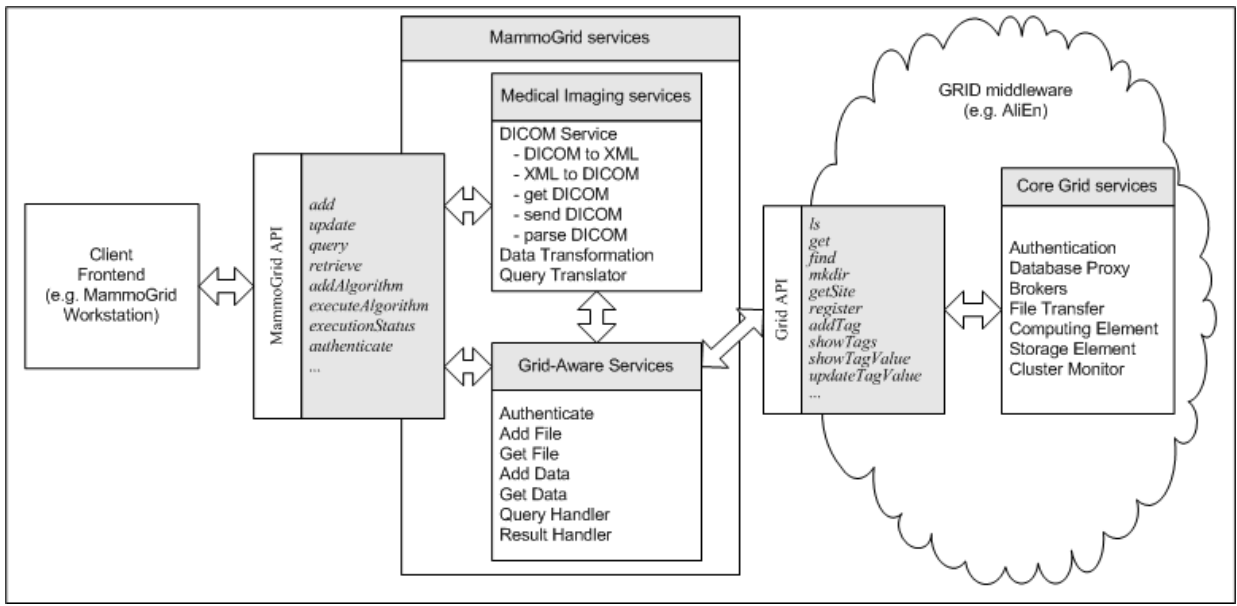

**Figure 1:** The high-level MammoGrid architecture

The MammoGrid architecture is Web Services (WS) based. Web Services address many of the lower-level needs of the SOA and are becoming increasingly common in the delivery of Grid-based solutions. Adopting a SOA therefore gives the MammoGrid the greatest future-proofing in terms of being able to cope with Grids interoperability and will, in principle, allow any Grid solution to interchange data with a Web-services based Grid.

**4.2. Handling interactions between clients and the Grid middleware**

Figure 2 diagrammatically represents one of the typical interactions between a medical client and the Grid – storing a new image in a DICOM file. DICOM (Digital Image Communication in Medicine) [10] is a de-facto standard in information exchange in medical imaging is and defines data structures for medical images and related data and transport protocol for communication between medical imaging equipments and other systems.

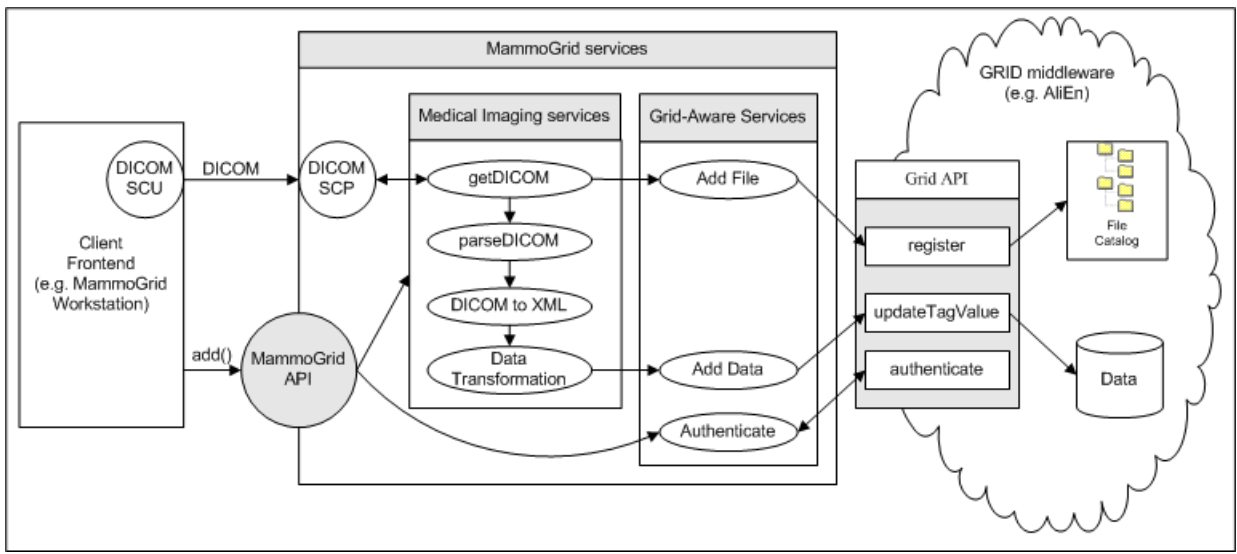

**Figure 2:**. Acquire a new image.

The MammoGrid project aims to conform to the DICOM standard in two ways. First, the digitized images should be imported and stored in the DICOM storage format (as DICOM files), so that the full set of image- and patient-related metadata is readily available with the images, and that information exchange with other medical devices understanding the DICOM storage format is seamless. To further ensure the compatibility with DICOM conformant clients, it is required that the exchange of DICOM datasets should be done via the communication protocol also defined by the standard. In this setup a client, the so-called 'Service Class User' (SCU), initiates a network connection with a server, or 'Service Class Provider' (SCP) and they exchange DICOM datasets over the established association protocol.

1. The MammoGrid service layer exposes a DICOM SCP which is capable of establishing an association with an SCU started by the MammoGrid workstation. Mammogram is transferred to the server via the SCU/SCP pair and stored on the server.
2. The client initiates processing of the DICOM file by calling the method add() of MammoGrid API. (This being a separate step from 1. is necessary as the DICOM protocol can not provide grid-aware authorization and authentication presently).
3. At MI services layer, associated image data such as image size, pixel values, resolution, and non-image data such as patient identification and demographic information, patient orientation, and acquisition parameters are extracted (parseDICOM), converted to generic XML representation (DICOM to XML) and transformed to the MammoGrid schema.
4. At the GA service layer, a DICOM file is saved in the Grid File Catalog (Add File) and the associated data are stored in the database.

As a practical example, the next section describes how clinical queries can be handled by the use of the MammoGrid Information Infrastructure.

## 5. Query Handling in MammoGrid

The MammoGrid project aims to provide a proof-of-concept demonstrator that allows clinicians to analyze mammograms resident across a Grid infrastructure. Clinicians define their mammogram analysis in terms of queries they wish to be resolved across the collection of data repositories. According the requirements, such queries can be categorized to simple queries (mainly against associated data stored in the database as simple attributes) and complex queries which require derived data to be interrogated or an algorithm to be executed on a (sub-)set of distributed images. The important aspect is that image and data distribution is transparent for radiologists so queries are formulated and executed as if these images were locally resident. The initial MammoGrid prototype (prototype P1) is based on a single Virtual Organization (VO) implementation of AliEn with simplified and centralized image metadata. Figure 3 illustrates the query handling in P1.

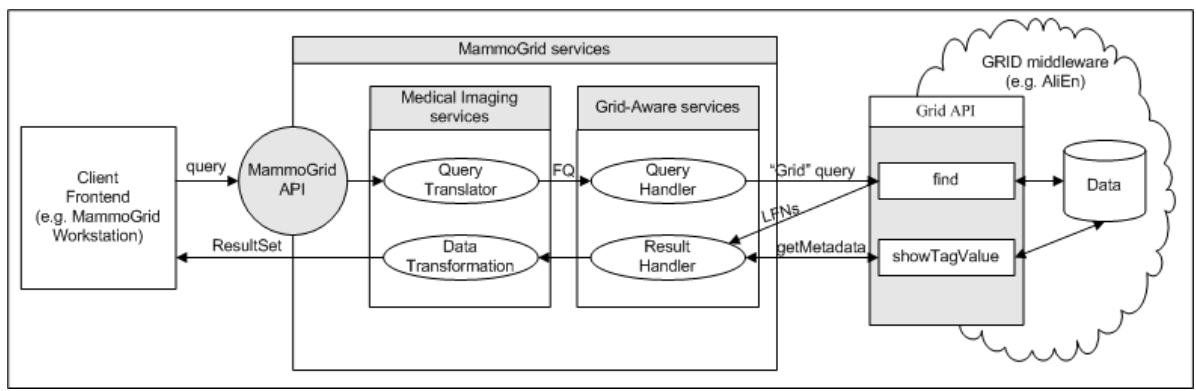

**Figure 3:** Query Handling in MammoGrid prototype P1.

Clients define their mammogram analysis in terms of queries they wish to be executed. The format of the query (e.g. XML) is passed to the *QueryTranslator* which is constrained by an agreement (e.g. XML Schema) between clients and MammoGrid service layer. There can be a number of such agreements between different clients and MammoGrid service layer, so the responsibility of the *QueryTranslator* is to translate the user request to the MammoGrid unified format (Formal Query, FQ), which is described by the XML Schema shown in Figure 4.

```xml
<?xml version="1.0" encoding="UTF-8"?>
  <xs:simpleType name="Comparison">
    <xs:restriction base="xs:string">
      <xs:enumeration value="EQUAL"/>
      <xs:enumeration value="LIKE"/>
      <xs:enumeration value="GREATER"/>
      <xs:enumeration value="GREATER_OR_EQUAL"/>
      <xs:enumeration value="SMALLER"/>
      <xs:enumeration value="SMALLER_OR_EQUAL"/>
      <xs:enumeration value="NOT_EQUAL"/>
    </xs:restriction>
  </xs:simpleType>

  <xs:simpleType name="Conjunction">
    <xs:restriction base="xs:string">
      <xs:enumeration value="and"/>
      <xs:enumeration value="or"/>
    </xs:restriction>
  </xs:simpleType>

  <xs:complexType name="Constraint">
    <xs:sequence>
      <xs:element name="Conjunction" type="Conjunction"/>
      <xs:element ref="pns:PatientModule"/>
      <xs:element name="Comparison" type="Comparison" minOccurs="0"/>
    </xs:sequence>
  </xs:complexType>

  <xs:element name="Query" type="ans:Query"/>
  <xs:complexType name="Query">
    <xs:sequence>
      <xs:element name="Constraint" type="Constraint" maxOccurs="unbounded"/>
      <xs:element name="QueryOrder" type="xs:string" minOccurs="0" maxOccurs="unbounded"/>
      <xs:element name="QueryLimit" type="xs:int" minOccurs="0"/>
      <xs:element name="QueryOffset" type="xs:int" minOccurs="0"/>
      <xs:element name="QueryNoData" type="xs:boolean" minOccurs="0"/>
    </xs:sequence>
  </xs:complexType>
</xs:schema>
```

**Figure 4:** Formal Query representation XML Schema.

A formal Query is passed to the GA service layer where the *QueryHandler* generates a query request in a form acceptable to the Grid middleware which then executes it using the Grid API. The Grid middleware carries out the request (find) and returns the list of Logical File Names (LFNs) of mammograms found in the file catalogue. The *ResultHandler* collects LFNs, gets their associated metadata and returns XML results to the client directly or, if needed, to the *Transform* MI service which is responsible for wrapping XML results in the format requested by the client (e.g. XML, WebRowSet etc.).

In the final stage of the project, the MammoGrid prototype (P2) will have distributed metadata. The query handling envisaged in P2 is presented in Figure 5.

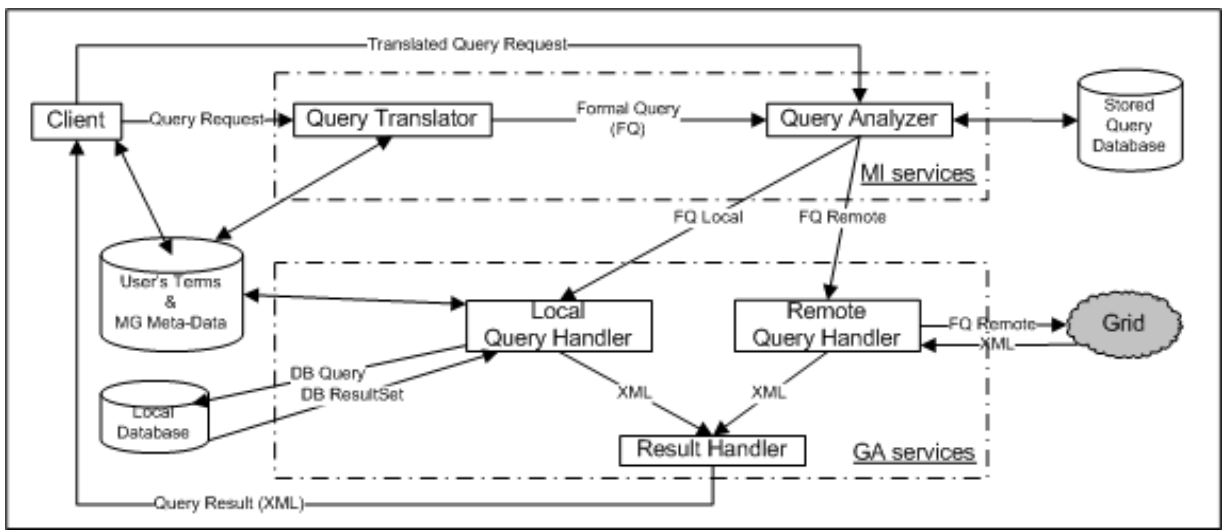

**Figure 5:** MammoGrid approach for handling medical queries.

In P2 queries are executed at the location where the relevant data resides, i.e. sub-queries are moved to the data, rather than large quantities of data being moved to the clinician, which can be prohibitively expensive given the quantities of data. The *Query Analyser* takes a formal query representation and decomposes into (a) formal query for local processing, and (b) formal query for remote processing. It then forwards these decomposed queries to the *Local Query Handler* and the *Remote Query Handler* for the resolution of the request. The *Local Query Handler* generates query language statements (e.g. SQL) in the query language of the associated Local DB (e.g. MySQL). The result set is converted to XML and routed to the Result Handler. The *Remote Query Handler* is a portal for propagating queries and results between sites. This handler forwards the formal query for remote processing to the Query Analyser of the remote site. The remote query result set is converted to XML and routed to the Result Handler.

Grid middleware can provide its own facilities for data access and integration. For example, data sources can be represented as Grid Data Services [6] that are managed by Grid middleware. In this case, Local Query Handler and Remote Query Handler can be used as mediators promoting loose coupling between medical imaging services and Grid middleware.

## 6. Conclusions

The MammoGrid project has recently delivered its first proof-of-concept prototype enabling clinicians to store (digital) mammograms along with appropriately anonymized patient meta-data and to provide controlled access to mammograms both locally and remotely stored. A typical database comprising several hundred mammograms is being created for user tests of the query handler. The prototype comprises a high-quality clinician visualization workstation [11] used for data acquisition and inspection, a DICOM-compliant interface to a set of medical services residing on a so-called 'Grid-box' and secure access to a network of other Grid-boxes connected through Grids middleware [12]. Clinicians are being closely involved with these tests and it is intended that a subset of the clinician queries listed in section 3 will be executed to solicit user feedback. Within the next year a rigorous evaluation of the prototype will then indicate the usefulness of the Grid as a platform for distributed mammogram analysis and in particular for resolving clinicians' queries. The system will be tuned for performance and for security prior to the release of a second prototype at the end of the project in mid 2005. It is intended that the MammoGrid medical services for this second prototype will adhere to emerging Grids standards.

The proliferation of information technology in medical practice and research will undoubtedly continue, addressing clinical demands and providing increasing functionality. The MammoGrid project is advancing deep inside this territory and exploring the requirements of evidence-based, computation-aided radiology, as specified by medical scientists and practicing clinicians. The very nature of a project like MammoGrid implies that it is inconceivable to define an exhaustive list or even complete classification of all possible queries which the radiologists may need to run against the distributed database. Inevitably, when the user community starts using such a system, the requirements will undergo adjustments and extension. This paper has illustrated the kind of complexity of the expected queries, based on an initial consultation of radiologists. It is proposed that the design followed, with extensive use of meta-data, is both capable of handling such complex queries in an efficient way and flexible enough to adapt to changing requirements. A design which handles queries using a reflexive data model has been presented as the proposed query model for the MammoGrid infrastructure.

In its first year, the MammoGrid project has faced interesting challenges originating from the interplay between medical and computer sciences and has witnessed the excitement of the user community whose expectations from the a new paradigm are understandably high. As the MammoGrid project moves into its final implementation and testing phase, further challenges are anticipated to every aspect of the project, from the definition of user requirements through to the very architecture of the system; this will test these ideas to the full.

## Acknowledgements

The authors take this opportunity to acknowledge the support of their institutes and their MammoGrid project colleagues for their contribution to different aspects of this research project.